\documentclass[aps,prl,twocolumn]{revtex4-2}
\usepackage{amsmath,amssymb,bm}
\begin{document}
\title{The Nonintrinsic Sector of Landau Theory}
\author{Trey Li}
\affiliation{University of Manchester, Manchester, UK}
\date{March 23, 2026}

\begin{abstract}

Landau theory usually treats free-energy coefficients as intrinsic parameters fixed by thermodynamic variables. We show that externally written microscale fields can survive coarse graining and enter the free-energy functional as spatially prescribed coefficient fields. This defines a nonintrinsic sector of Landau theory. The key condition is a hierarchy of correlation, writing, and frustration lengths. We identify ion-patterned FeRh as a plausible realization.

\end{abstract}
\maketitle

Landau theory gives a standard description of phases and their transitions \cite{Landau1937,Goldenfeld1992,Kardar2007}. It encodes order-parameter physics through a continuum free-energy functional of the form
\begin{equation*}
F[m]=\int d\mathbf r\,
f\!\left(m(\mathbf r),\nabla m(\mathbf r);\{\lambda_i\}\right).
\end{equation*}
Here $m$ is the order parameter, $\mathbf r$ the position, and ${\lambda_i}$ are coefficients such as the quadratic $a$, quartic $b$, and gradient stiffness $\kappa$. In conventional formulations, these coefficients are intrinsic. They are inherited from the microscopic Hamiltonian and fixed by global variables such as temperature, stress, and composition. The Landau functional is therefore an inherited object and not a degree of freedom.

This intrinsic assumption is usually taken for granted. Yet some modern materials allow nanoscale control of defects, disorder, or local fields \cite{Heidarian2015,Koide2014,Cress2021}. Such control can modify local phase preference. A natural question is whether this control extends to the Landau coefficients $\lambda_i(\mathbf r)$. If so, the resulting coefficient fields are not simply forms of disorder \cite{Nishimori2011} or inherited heterogeneity \cite{Chen2002,Milton2002}, but externally assigned inputs to the Landau functional.

To formulate this, consider a field $D(\mathbf r)$ externally written into a material at the microscale.
It couples locally to the order parameter $m$ in the microscopic Hamiltonian
\begin{equation*}
H = H_0[m] + \int d\mathbf r\, V\!\left(m(\mathbf r), D(\mathbf r)\right).
\end{equation*}
Here $H_0$ encodes the intrinsic interactions, and $V$ encodes the effect of $D$.
Let $\ell_D$ be the variation scale of $D(\mathbf r)$.
We assume that $D$ can be patterned at this resolution by techniques such as ion irradiation \cite{Koide2014,Heidarian2015,Cress2021} or local compositional doping \cite{LeGraet2015}. The question is whether such an imposed pattern can persist as a coefficient field in the Landau functional.

We identify a simple criterion. Let $D(\mathbf r)$ be a field written at the microscale.
If its variation scale exceeds the correlation length, the imposed pattern survives coarse graining \cite{Sethna2006,Presutti2009}.
If it also remains inert to long-range reconstruction, it appears in the Landau functional as a stable coefficient field.

For the first condition, the idea is that large-scale structure survives small-scale averaging.
To see this, recall how coarse graining works. Let $\xi$ be the correlation length of the order parameter $m$.
Coarse graining integrates out fluctuations on scales $\lesssim \xi$ \cite{Sethna2006}.
To leading order in a gradient expansion, the coarse-grained free energy retains the Landau form
\begin{equation*}
F[m]=\int d\mathbf r\,
\left[a_{\rm eff}(\mathbf r)m^2
+ b_{\rm eff}m^4 + \frac{\kappa_{\rm eff}}{2}|\nabla m|^2+ \cdots\right].
\end{equation*}
Only the coefficients are renormalized. They are set by microscopic couplings, including the presence of $D(\mathbf r)$.
In particular, $D$ enters the quadratic term, with its effect captured by a local mapping
\begin{equation}
a_{\rm eff}(\mathbf r)
  = \mathcal A[D(\mathbf r)]
  + O\!\left(\nabla^2 D,\,|\nabla D|^2\right).
\label{eq:aeff_local}
\end{equation}
Here $\mathcal A$ is a smooth function fixed by microscopic interactions.
The gradient corrections are higher order. They are small provided $D(\mathbf r)$ varies slowly on the scale of $\xi$.
They can therefore be neglected at leading order \cite{Presutti2009}.

To make this more explicit, write the coarse-grained coefficient in convolution form
\begin{equation*}
a_{\rm eff}(\mathbf r)
  = \int d\mathbf r'\, K_\xi(\mathbf r-\mathbf r')
    \,\mathcal A[D(\mathbf r')].
\end{equation*}
This expresses the coarse-grained coefficient as a local spatial average of the mapping $\mathcal A[D]$.
The kernel $K_\xi$ is short-ranged, with width set by $\xi$. Thus only nearby points contribute.
When the variation scale $\ell_D$ is much larger than $\xi$, this averaging does not significantly smear the written pattern.
The coefficient $a_{\rm eff}(\mathbf r)$ then follows $\mathcal A[D(\mathbf r)]$ closely. This yields the first condition:
\begin{equation}
\xi \ll \ell_D.
\label{eq:cond_xi}
\end{equation}
When Inequality~\eqref{eq:cond_xi} holds, the imposed field $D(\mathbf r)$ survives coarse graining.

For the second condition, the idea is that small-scale structure resists long-range reconstruction. Relevant long-range mechanisms may include elastic frustration, dipolar interactions, and electrostatic effects. Elastic frustration provides one concrete example of how long-range back-action can generate a competing reconstruction length scale \cite{Paez2016}, while general long-range couplings are standard in continuum condensed-matter descriptions \cite{Chaikin1995}. Let $\ell_{\rm fr}$ be the frustration length associated with the dominant long-range back-action. If the imposed scale $\ell_D$ exceeds this length, the pattern is reconstructed by internal forces. A written pattern is stable only when $\ell_D$ lies well below the frustration length. This yields the second condition:
\begin{equation}
\ell_D \ll \ell_{\rm fr}.
\label{eq:cond_fr}
\end{equation}
When Inequality~\eqref{eq:cond_fr} holds, $D(\mathbf r)$ remains metastable and inert on experimental time scales.

Combining Inequalities~\eqref{eq:cond_xi} and \eqref{eq:cond_fr}, we arrive at the scale hierarchy
\begin{equation}
\xi \ll \ell_D \ll \ell_{\rm fr}.
\label{eq:hierarchy}
\end{equation}
When Inequality~\eqref{eq:hierarchy} holds, the Landau functional acquires coefficient fields $\lambda_i(\mathbf r)=\lambda_{i,\rm eff}(\mathbf r)$ that are determined not by global thermodynamic variables but by the imposed patterns $D_i(\mathbf r)$. This scale hierarchy therefore gives a simple criterion for the nonintrinsic sector and defines the intrinsic and nonintrinsic sectors of Landau theory:
\begin{align*}
\text{intrinsic: }& \lambda_{i,\rm int}=\lambda_i(T,\varepsilon,\mu,\dots),\\
\text{nonintrinsic: }& \lambda_{i,\rm nint}=\lambda_i(\mathbf r)\ \text{externally prescribable}.
\end{align*}
In the intrinsic sector, coefficients are fixed by global variables. In the nonintrinsic sector, they are writable fields. 

Two familiar sources of spatial coefficient variation lie outside the nonintrinsic regime. In quenched-disorder settings, the field is stochastic and typically renormalizes into a disordered but still intrinsic coefficient \cite{Nishimori2011}. In heterogeneous, composite, or otherwise microstructure-derived systems, coefficient variation may be deterministic, but it is fixed by the material architecture itself and is not freely rewritable during operation \cite{Chen2002,Milton2002}. By contrast, in the nonintrinsic regime the coefficient field is neither stochastic nor merely inherited, but externally assigned and thermodynamically inert. It therefore defines a distinct class of physical input to the Landau functional. 

In the nonintrinsic sector, Landau theory retains the algebraic form of the Landau expansion and the standard continuum dynamics,
\[
\partial_t m=-\Gamma\,\frac{\delta F}{\delta m}+\eta.
\]
The gradient-flow structure is unchanged, and noise enters as usual \cite{HohenbergHalperin1977}. What changes is the geometry of the free-energy landscape, with direct dynamical consequences.
We first explain how the accessible family of landscapes is enlarged \cite{Toledano1987}, and then discuss the resulting dynamics.

In intrinsic systems, the coefficients $\lambda_{i,\rm int}=\lambda_i(T,\varepsilon,\mu,\dots)$ depend only on a few global variables such as $T,\varepsilon,\mu$. Spatial variation arises only indirectly, through gradients of these variables \cite{Chaikin1995}. As a result, the accessible family of Landau functionals is controlled by only $\mathcal O(1)$ global parameters. 

In contrast, in nonintrinsic systems the coefficient fields $\lambda_i(\mathbf r)$ can, in principle, be specified across the sample subject to the coarse-graining and frustration constraints. A natural measure of this enlargement is the effective number of independently writable coarse-graining regions. In the idealized limit in which $\ell_D$ is the dominant resolution scale and long-range constraints do not correlate distant regions, one obtains the scaling estimate
\begin{equation*}
\mathcal N_{\rm wrt}
  \sim \left(\frac{L}{\ell_D}\right)^d,
\end{equation*}
for that number. In practice, convolution at scale $\xi$ and long-range back-action at scale $\ell_{\rm fr}$ reduce the effective independent count. But the key point is that, in the nonintrinsic sector, the number of independent degrees of freedom can grow with system size, rather than being limited to a few global control variables.

To illustrate the geometric freedom, consider a one-dimensional system of length $L$, with $\xi \ll L$.
In the intrinsic regime, the coefficient $a$ is tuned through a global parameter, such as temperature.
The local potential shifts uniformly in space, and no spatial structure is introduced.
By contrast, in the nonintrinsic sector, spatial patterns can be encoded, for example as a periodic modulation,
\begin{equation*}
a(x)=a_0 + \delta a
      \cos\!\left(\frac{2\pi x}{\Lambda}\right),
\quad
\xi \ll \Lambda \ll \ell_{\rm fr},
\end{equation*}
or as more general landscapes generated by superposed localized modulations. Such patterns modify the local free-energy density, reshaping the landscape point by point and thereby steering the gradient-flow dynamics.

To see the local effect of such patterns, consider the local free-energy density
\begin{equation*}
f_{\rm loc}(m,x)=a(x)m^2 + b\,m^4,
\end{equation*}
with $b>0$. It depends on the order parameter $m$ and the position $x$.
The positive quartic term ensures stability.
The coefficient $a(x)$ varies in space and can be interpreted as encoding the local preference for antiferromagnetic (AF) or ferromagnetic (FM) order in systems such as FeRh \cite{Staunton2014}.
Regions with $a<0$ exhibit a double-well potential, while regions with $a>0$ exhibit a single-well potential.
A prescribed $a(x)$ thus produces a spatial pattern of bistability, locally favoring or suppressing order.

As for the consequences for gradient-flow dynamics, we continue to use the quadratic coefficient $a(\mathbf r)$ as an example.
Under purely relaxational dynamics, the free energy decreases monotonically according to
\begin{equation*}
\frac{dF}{dt}
  = -\Gamma \int\!\left|\frac{\delta F}{\delta m}\right|^2 d\mathbf r
  \le 0.
\end{equation*}
This is the standard dissipation relation for relaxational dynamics \cite{HohenbergHalperin1977}.
When domain walls are present, curvature drives their motion, while spatial variations in $a(\mathbf r)$ bias it.
Schematically, one may write
\begin{equation}
v_n=-\Gamma\left(\sigma\kappa_{\rm geom}
 + \partial_n a\,\Delta m^2\right),
\label{eq:v_n}
\end{equation}
where $v_n$ is the normal velocity, $\sigma$ the interfacial tension, $\kappa_{\rm geom}$ the geometric curvature, and $\Delta m$ is the jump in the order parameter across the interface \cite{Allen1979,Chen2002}.

Equation~\eqref{eq:v_n} shows a dynamical consequence of the nonintrinsic sector: gradients in a written coefficient field can bias domain-wall motion without requiring a time-dependent external drive during operation. This is not random pinning from quenched disorder, but a directed bias set by the prescribed geometry of the free-energy functional. This mechanism may also suggest computational uses of prescribed free-energy landscapes, a direction we do not pursue here. 

A plausible realization of the nonintrinsic Landau sector is provided by the metamagnetic alloy FeRh.
Near the transition, the relevant correlation length is short compared with the patterning scale, while ion irradiation enables the writing of chemical-disorder patterns on length scales of tens of nanometers \cite{Cress2021}, consistent with the first condition $\xi \ll \ell_D$.
The AF--FM transition is weakly first order and accompanied by only a small volume change of order $1$--$2\%$ \cite{Cooke2012}.
This is consistent with relatively weak long-range elastic frustration over experimentally relevant pattern scales, supporting $\ell_D \ll \ell_{\rm fr}$ for appropriate write protocols.
The hierarchy $\xi \ll \ell_D \ll \ell_{\rm fr}$ can therefore be satisfied in FeRh for appropriate write protocols and pattern scales.

FeRh is a particularly promising case because its AF--FM transition is highly sensitive to chemical disorder and composition \cite{Staunton2014}.
This suggests that a written disorder field $D(\mathbf r)$ can map smoothly onto the Landau coefficient $a(\mathbf r)\propto T-T_{\mathrm t}(\mathbf r)$.
More specifically, ion irradiation introduces local chemical disorder, and the transition temperature shifts systematically with irradiation dose \cite{Cress2021,Heidarian2015}.
Let $D(\mathbf r)$ be a local disorder variable introduced at spatial resolution $\ell_D$.
A local expansion of the AF--FM free-energy difference may then be written as
\begin{equation*}
\Delta f(D) \approx
\alpha_1 D + \alpha_2 D^2 + \cdots,
\end{equation*}
where $\alpha_{1},\alpha_{2}$ are material-specific constants determined phenomenologically or from first-principles calculations.
Since $a(\mathbf r)\propto T-T_{\mathrm t}(\mathbf r)$ and $T_{\mathrm t}$ is disorder-sensitive, this gives a smooth mapping $a(\mathbf r)=\mathcal A[D(\mathbf r)]$, consistent with Equation~\eqref{eq:aeff_local}.
Such a mapping remains effectively local on the coarse-graining scale provided defect diffusion is negligible and magnetostructural back-action does not reconstruct the written pattern.

Experiments already demonstrate the relevant controllable spatial modulations in FeRh.
Controlled AF--FM interface motion has been achieved using deliberately imposed transition-temperature gradients, for example through dopant-graded growth \cite{LeGraet2015}.
Direct-write nanoscale patterning and dose-tunable suppression of the transition temperature using focused He$^+$ irradiation have also been demonstrated \cite{Cress2021}.
Available experiments further indicate that spatially patterned transition behavior and phase coexistence can persist over experimentally relevant windows and correlate with reproducible spatial variations in local phase behavior \cite{Cress2021,Griggs2020}. In the coexistence regime, such written landscapes bias AF--FM boundary motion in the direction expected from Equation~\eqref{eq:v_n}, provided the pattern is not erased by annealing or reconstructed by long-range elasticity \cite{LeGraet2015}.

Taken together, these observations make FeRh a candidate platform for exploring the nonintrinsic sector of Landau theory. They show that controllable spatial modulations can be written, retained over an operating window, and translated into reproducible biases of relaxational dynamics.
The resulting evolution remains relaxational, but proceeds along a prescribed landscape rather than one inherited solely from global thermodynamic variables. In this sense, the framework does not replace the phenomenology, but organizes it in terms of coefficient-field writability. 

Not all materials permit externally writable coefficient fields.
We now identify boundaries of the nonintrinsic sector and the mechanisms that exclude materials from it.

A first feature is large transformation strain. In martensitic and ferroelastic systems, phase coexistence induces long-range elastic compatibility stresses.
These stresses can reconstruct local phase patterns and drive the written field $D(\mathbf r)$ toward equilibrium, reducing the range of scales over which independent writability can be maintained.
The resulting $\ell_{\rm fr}$ can then become microscopic, violating $\ell_D \ll \ell_{\rm fr}$ and excluding stable writability.

A second feature is strong dipolar or magnetostatic coupling. In ferromagnets and ferroelectrics with dominant long-range fields, variations in $D(\mathbf r)$ can be overridden by magnetostatic or electrostatic interactions.
The effective free-energy landscape then becomes strongly constrained by the nonlocal interaction kernel, and coefficient shaping beyond that kernel is limited.

A third feature is mobility of the written field. If $D(\mathbf r)$ relaxes by diffusion, defect aggregation, or charge migration, it cannot act as a fixed input to the Landau functional.
Fields that equilibrate under mild annealing or screening therefore lie outside the nonintrinsic regime.

A fourth feature is strong electronic or structural back-reaction. When variations in $D(\mathbf r)$ substantially modify the electronic structure or induce phase separation, the map $D \mapsto a$ becomes nonlocal.
The locality assumed in Equation~\eqref{eq:aeff_local} then fails, and the effective coefficient field ceases to be independently writable.

Materials exhibiting any of these mechanisms cannot support stable externally prescribable coefficient fields, even when nanoscale patterning is available.

In conclusion, we have identified a nonintrinsic sector of Landau theory and shown that externally written microscale fields can survive coarse graining and enter the free-energy functional as coefficient fields. In this sector, part of the functional is no longer inherited solely from global thermodynamic variables, but can be externally prescribed.

The central condition is the scale hierarchy in Equation~\eqref{eq:hierarchy}. FeRh provides a plausible realization of this regime, while the exclusion mechanisms identified here clarify why such writability should not be expected generically across phase-transition materials. The conceptual shift is simple: Landau theory need not describe only inherited free-energy landscapes. In the nonintrinsic sector, it also admits written ones.

The present work focuses on spatially writable coefficient fields. Whether the nonintrinsic sector admits more general forms of external prescription remains an open question.

\end{document}